\documentclass[12pt,preprint]{aastex}

\shorttitle{Moments Method}
\shortauthors{Estrada and Cuzzi}

\begin{document}

\title{Solving the Coagulation Equation by the Moments Method}

\author{P. R. Estrada}
\affil{SETI Institute, 515 N. Whisman Rd., Mountain View, CA 94043}

\and

\author{J. N. Cuzzi}
\affil{NASA Ames Research Center, MS 245-3, Moffett Field, CA 94035}

\begin{abstract}
We demonstrate an approach to solving the coagulation equation that involves
using a finite number of moments of the particle size distribution. This 
approach is particularly useful when only 
general properties of the distribution, and their time evolution, are needed. 
The numerical solution to the integro-differential Smoluchowski coagulation
equation at every time step, for every particle size, and at every spatial 
location is
computationally expensive, and serves as the primary bottleneck in running
evolutionary models over long periods of time. The advantage of using 
the 
moments method comes in the computational time savings gained from only
tracking the time rate of change of the moments, as opposed to tracking the 
entire mass histogram which can contain hundreds or thousands of
bins depending on the desired accuracy. The collision kernels of the
coagulation equation contain all the necessary information about particle
relative velocities, cross-sections, and sticking coefficients. We show how
arbitrary collision kernels may be treated. We discuss particle relative
velocities in both turbulent and non-turbulent regimes.
We present examples of this approach
that utilize different collision kernels and find good agreement between
the moment solutions and the moments as calculated from direct integration
of the coagulation equation. As practical applications, we demonstrate
how the moments method can be used to track the evolving opacity, and
also indicate how one may incorporate porous particles.
\end{abstract}


\section{Introduction}

The conditions under which planetary formation is initiated in the 
protoplanetary disk remain poorly understood despite several decades of 
astrophysical research. Furthermore, the continued 
discovery of extrasolar planets (over two hundred currently) further 
emphasizes the need to understand the formation of not only our own solar 
system, but of these numerous strange and diverse systems as well. 
At the forefront still remains the basic question of 
how growth occurs from dust to planet. The key properties of protoplanetary 
nebulae remain controversial, yet grain growth from dust to larger 
agglomerates 
($\sim $ cm-m), and from planetesimals ($\gtrsim$ km) to planets must have 
occurred in some manner. While numerous studies of $N$-body dynamics have 
addressed the growth of terrestrial planets from asteroid-size planetesimals
which are largely decoupled from the gas \citep[e.g.,][]{lei05,bro06},
it is the transition from agglomerates to 
planetesimals which continues to provide the major stumbling block in 
planetary origins \citep{wei97,wei02,wei04,wei93,ste97,dul05,cuz06,dom07}.

Grain growth begins with sticking of sub-micron sized grains, which are 
dynamically coupled to the nebula gas and collide at
low relative velocities which are 
size-dependent \citep{wei93}. These relative velocities can be caused
by a variety of mechanisms, such as Brownian motion for the smallest grains,
differences in coupling to eddies in a
turbulent nebula \citep{vol80,wei84,mar91,wei93,cuz03,orm07}, 
and/or vertical, radial, and azimuthal drift driven by global gas pressure 
gradients
\citep{nak86,wei97}. For the 
smallest sized grains and aggregates, sticking is caused by weak van der 
Waals interaction forces \citep[although stronger forces may also act; see, 
e.g.,][]{dom07}. This sticking forms larger and larger aggregates
until particles grow large enough to decouple from the gas and
settle to the midplane where they may, barring strong turbulence
\citep[although see][for an exception]{joh07}, grow into
larger objects, even planetesimals \citep{saf69,wei93,cuz93,wei97}.

There is considerable observational evidence supporting dust growth 
in protoplanetary disks, at least from sub-micron to millimeter scales, based
on observations of the dust continuum emission originating from protoplanetary 
disks. The wavelength dependence of dust emission from IR to radio wavelengths
suggests, in many cases, that typical grain sizes are in the millimeter range,
which is orders of magnitude larger than interstellar grains
\citep{bec91,dul05,dom07}. 
Measurements of the contrast of the Orion trapezium 
silicate emission feature at 10 $\mu$m relative to the local continuum implies 
an increase in size from interstellar dust (sub-micron) to micron sizes in the
disk photospheres \citep{van03,pry03,kes06}.

Theoretical approaches to the study of dust coagulation normally involve 
solving the cumbersome collisional coagulation equation (see Eq. 1) at each
spatial location ($R,\phi,Z$) in the disk. 
The collisional kernel $K(m,m^\prime)$ for particles of mass $m$ and $m^\prime$
usually involves some sort of sticking efficiency, and particle cross-sectional
area, in addition to the particle-particle 
relative velocities. Although the {\it systematic} particle-particle
velocities that arise from 
gas pressure gradients \citep{nak86} are 
somewhat easy to incorporate into the collisional kernel, 
relative velocities in turbulence are more complicated.
If one assumes that nebula turbulence follows a Kolmogorov inertial range
\citep{vol80} in which energy
cascades from the largest, slowly rotating eddies, down to some minimum
lengthscale where the intrinsic molecular viscosity can dissipate the
macroscopic gas motions (and turbulence ceases), one can obtain closed-form
expressions for the particle-particle, and particle-gas
turbulent relative velocities \citep{cuz03}, even for particles
of different sizes \citep{orm07}. Numerical
models of particle growth in the inner regions of protoplanetary disks 
indicate that inclusion of these turbulence-induced relative 
velocities is essential \citep{dul05}, as it provides a
source of energetic collisions that, in the presence of fragmentation
and destruction, leads to the
ongoing existence of small grains even after several million years.

A full scale solution to the problem of dust coagulation for a size spectrum 
at every vertical and  radial location in the protoplanetary disk, as a
function of time,
is thus prohibitive even with today's computational capabilities
\citep{wei02,wei04}. So it is
advantageous to find alternative methods of extracting information from the
coagulation process, such as time scales of growth, the largest particle 
size, the size that dominates the area (radiative transfer), and the size 
that dominates
the mass (migration and redistribution) without having to track the behavior 
of the entire mass distribution. 

In
this paper, we describe a method that utilizes the moments of the coagulation
equation to accomplish this task, in particular when
the functional form of the mass distribution can be assumed. In {\S} 2, 
we derive the
moment equations from the coagulation equation, and compare
direct integration of the coagulation equation with solutions using
the moments approach in the case of a simple turbulent coagulation kernel. 
We also present  
two different approaches to obtaining solutions using a realistic kernel,
when the form of the size distribution is assumed to be a powerlaw. 
In {\S} 3, we evaluate the different approaches developed in {\S} 2
for realistic kernels by numerical integration of the moments equations
and compare these moments to those obtained by direct integration of the
coagulation equation.
We also
compare the moments solution to an alternate analytical approach by
\citet{gar07}. In {\S} 4, as a demonstration of the practical application 
of 
the moments method, we calculate the wavelength-independent opacity, and
discuss generalization to wavelength-dependent Planck or Rosseland
mean opacities. In {\S} 5, we indicate how porosity may be included.
In {\S} 6, we present our conclusions. 

\section{Solution Techniques Using the Moments Method}

The moments method approach to modeling particle growth allows an attractive
alternative to solving the full coagulation equation when it is only necessary
to know a particular property, or properties of the evolving particle size
distribution. The coagulation equation \citep{smo16} in its
integro-differential form is given by

\begin{eqnarray}
\frac{df(m,t)}{dt} = \frac{1}{2}\int_{0}^{m} K(m-m^{\prime},
m^{\prime}) f(m-m^{\prime},t) f(m^{\prime},t) dm^{\prime} - \nonumber \\
\int_{0}^{\infty}
K(m,m^{\prime})f(m,t) f(m^{\prime},t) dm^{\prime} \,\,\,+ \,\,\,
\frac{df}{dt}\Big|_{+}\,\,\,+\,\,\, \frac{df}{dt}\Big|_{-}\,\,,
\end{eqnarray}

\noindent
where $f(m,t)$ is the particle number density per unit mass at mass $m$, and
$K(m,m^{\prime})$ is the collision kernel connecting the properties of 
interacting
masses $m$ and $m^\prime$, which typically takes into account mutual particle
cross section $\sigma(m,m^\prime)$, relative velocity $\Delta V(m,m^\prime)$, 
and a particle sticking efficiency $S(m,m^\prime)$. The last two terms on
the RHS of Eq. (1) represent sources and sinks such as particle
erosion, fragmentation and subsequent redistribution of the fragmented
population, and gravitational growth. Although we will not specifically
address these mechanisms, we will indicate how they may be treated, and
save implementation for a forthcoming paper.

The
motivation behind using the moments method is to avoid the computational cost
inherent in Eq. (1) which entails solving the convolution integral (the first
term on the RHS of Eq. 1) at every location, and at every time step. 
Depending on the functional form of the kernel, the second integral on the
RHS may also need to be integrated repeatedly. Given
that the typical mass spectrum may involve $10^2-10^3$ bins to acquire the
desired accuracy, the computational burden required becomes a detriment to any
study involving a wide range of parameter space. The
so-called brute force solution of the coagulation equation thus becomes the 
primary bottleneck in running 2D evolutionary models over extended periods of 
time \citep{wei04,dul05}.

We define the $p$-th moment $M_p$ of the distribution, where $p$ need not be
an integer, as

\begin{equation}
M_p = \int_{0}^{\infty} m^p f(m,t) dm,
\end{equation}

\noindent
where the units of $f(m,t)$ are such that $M_0=\int f(m,t)\,dm$ represents 
the total number
density of particles, and $M_1 = \int mf(m,t)\, dm = \rho$ is the 
total volume mass density
of solids. The essence of the moments method is as follows. We multiply both
sides of the coagulation equation (Eq. 1) by $m^k$, where $k$ is an integer,
and then integrate both sides over mass $m$:

\begin{eqnarray}
\frac{dM_k}{dt} = \int_{0}^{\infty} m^k \frac{df}{dt} dm = 
\frac{1}{2} \int_{0}^{\infty} m^k\,dm \int_{0}^{m} K(m-m^{\prime},
m^{\prime}) f(m-m^{\prime},t) f(m^{\prime},t) dm^{\prime}  - \nonumber \\
\int_{0}^{\infty} m^k f(m,t)\,dm \int_{0}^{\infty}
K(m,m^{\prime}) f(m^{\prime},t) dm^{\prime}.
\end{eqnarray}

\noindent
We introduce a step function $H(m-m^\prime)$, such that $H = 0$ for
$m -m^\prime < 0$ and $H = 1$ otherwise, to extend the limits of the
integral over $m^\prime$ from ($0,m$) to ($0,\infty$). The convolution
integral (the first integral on the RHS of Eq. 3) then becomes

\begin{eqnarray}
\frac{1}{2}\int_{0}^{\infty} m^k\,dm \int_{0}^{\infty} 
H(m-m^\prime) K(m-m^\prime,m^\prime) f(m-m^\prime,t) f(m^\prime,t)\,
dm^\prime  = \nonumber \\
\frac{1}{2} \int_{0}^{\infty} f(m^\prime,t)\,dm^\prime \int_{0}^{\infty}
(u+m^\prime)^k K(u,m^\prime) f(u,t)\,du,
\end{eqnarray}

\noindent
where on the RHS of Eq. (4) we have switched the order of integration and
made the substitution $u = m - m^\prime$. The RHS side of Eq. (4) may then be
combined with the last double integral in Eq. (3) to yield the set of 
ordinary differential equations (ODEs)
for the integer moments \citep{mar01}:

\begin{equation}
\frac{dM_k}{dt} = \int_{0}^{\infty} \int_{0}^{\infty} \left[\frac{1}{2}
(m + m^\prime)^k - m^k\right] K(m,m^\prime) f(m,t) f(m^\prime,t) dm dm^\prime,
\end{equation}

\noindent
where we have substituted $m$ for $u$ with no loss of generality.
Depending on the mass dependence of the kernel (as illustrated below) 
the right hand side can readily be
expressed as products of moments of order $k$ or less, leading to a closed
system of equations which may be solved using standard techniques. 
In this definition of the coagulation equation in which it is assumed there
are no sources and sinks, the first 
moment $M_1 \equiv \rho$ is constant in time, that is 
$dM_1/dt = 0$\footnote{We note that under realistic protoplanetary nebula 
conditions, $\rho$ will not be constant due to, e.g., size-dependent advection
terms in the equations. Such effects can be treated separately from the
``coagulation'' step. See {\S} 2.2.2.}. 

Exact solutions to the coagulation equation have been obtained for 
some specific choices
of the kernel \citep[e.g.,][]{smo16,tru71}, the most simple 
being that of constant kernel $K(m,m^\prime) = \beta_0$. Although the exact
solution for $f(m,t)$ cannot be obtained from the moments equations, the 
time rate of
change of the zeroth moment ($k=0$), can easily be obtained from Eq. (5)
which reduces to $dM_0/dt = - (1/2)\beta_0M_0^2$ and has the trivial solution
\citep[see, e.g.,][]{sil79}

\begin{equation}
M_0(t) = \frac{M_0(0)}{1 + (1/2)\beta_0M_0(0)t},
\end{equation}

\noindent
independent of the initial choice of distribution $f(m,0)$. Likewise, the
ODE for $M_2$, which can be associated with the density-weighted mean particle
size ($\left<m\right> = M_2/M_1$), yields the simple solution 
$M_2(t)=M_2(0) + \beta_0\rho^2t$. Despite not
knowing the exact solution, we are able
to understand the behavior of general properties of the
mass distribution with time through the moments equation without
tracking the behavior of the full mass spectrum. Thus, 
if it is only desired to know, for example, the time evolution of the 
particle representing most of the {\it area} (first moment) or most of
the {\it mass} (second moment) of the distribution, then the advantage of 
the moments method becomes clear. A small
number of moments is all that is necessary to determine the behavior of the
system. In particular, we will be interested in the size of the largest
particle $m_L(t)$ in the entire mass distribution as a function of time
(which we show below can be computed from the integer moments),
because these are usually the most rapidly drifting and most violently
colliding particles \citep{cuz06}.

\subsection{Example: Saffman and Turner Turbulent Coagulation Kernel}

We can illustrate the moments method approach using a very simple turbulent 
coagulation kernel
where $K(m,m^\prime) = \gamma_0 (m^{1/3} + m^{\prime 1/3})^3$ \citep{saf56},
and no sources or sinks. Physically, this represents the 
product of $(r + r^{\prime})^2$ for
area, and $(r + r^{\prime})$ for relative velocity of two particles in a 
laminar shear flowfield. If a sticking coefficient were desired then we
would specifically have $K(m,m^\prime) = \gamma_0(m^{1/3}+m^{\prime 1/3})^3
S(m,m^\prime)$ where $0\leq S(m,m^\prime) \leq 1$. Here, $\gamma_0$ is a 
constant that depends on the Reynolds number of the gas flow.
Inserting this kernel into Eq. (5), we find the set of equations

\begin{eqnarray}
\frac{dM_0}{dt} = -\gamma_0(M_0M_1 + 3M_{1/3}M_{2/3}), \,\,\,\,\,
\frac{dM_1}{dt} = 0 \nonumber \\
\frac{dM_2}{dt} = 2\gamma_0(M_2M_1 + 3M_{4/3}M_{5/3}). 
\end{eqnarray}

\noindent
We note that the physical derivation of kernels in terms of particle radius $r$
leads to {\it fractional}
moments in terms of $m$. These fractional moments look complicated, but can 
be solved for by simple
interpolation using Lagrange polynomials in terms of the more familiar integer
moments \citep{log79,pre92,mar01}.
Thus, any
fractional moment $M_p$ may be expressed compactly in the normalized form
$M^\prime_p(t) = M_p(t)/M_p(0)$:

\begin{equation}
M^\prime_p(t) = \prod_{j=k}^{k+n} \left[M^\prime_j(t)\right]^{L^n_j(p)},
\end{equation}

\noindent
where $n$ is the number of integer moments, $k = 0,1,...,n$,
and the exponent $L^n_j$ is defined as

\begin{equation}
L^n_j(p) = \frac{1}{n!}{\prod}_{n+1}(p)\frac{(-1)^{n-j}C^j_n}{p-j},
\end{equation}

\noindent
with ${\prod}_{n+1}(p) = p(p-1)...(p-n)$, and the $C^j_n$ 
are the binomial coefficients $n!/j!(n-j)!$. The 
important thing to
note here is that the order of the moments must remain less than or equal 
to the largest moment in order for the system to remain closed. In general,
this will be true for realistic collision kernels (see {\S} 2.2).

To test the accuracy of the moments method, we integrated equations (7) using a
fourth order Runge-Kutta scheme, and compared the results to a brute force
integration of the coagulation equation (Eq. 1). For the latter, we 
integrated the distribution function $f(m,t)$ at each timestep
to determine the moments of the distribution $M_0$, $M_1$, and $M_2$. We chose
$f(m,0) = c_0m^{-q}$ as our initial distribution for simplicity. Since the
form of the mass distribution is only assumed at $t=0$, we consider this to
be an example of an implicit assumption (see {\S} 2.2.2). 
The results of this calculation
are presented in Figure 1. We have used two different resolutions for the
brute force calculation in order to demonstrate how the higher resolution
(and much more computationally expensive) case (solid symbols) approaches the
moments approach solution. Notice that the first moment $M_1 = \rho$ remains
constant as expected. Given that
general numerical errors can arise from the finite mass grid and coarse 
timesteps ($\Delta t = 10$ years) used for 
the calculation, and that systematic errors may be 
introduced by the Lagrangian interpolation, the fit is quite good. 
It is important 
to point out here that while
the coagulation calculation required as many as $20-30$ hours of CPU time to 
complete on a 2 GHz machine, the moments calculation of Eq. (7) is 
essentially instantaneous.

\subsection{Realistic Coagulation Kernels}

The Saffman-Turner turbulent coagulation kernel we used as an example in 
{\S} 2.1
is simple to utilize and is expressible explicitly in powers of the mass $m$.
In practice, however, the realistic coagulation kernels that we will be using
will be more complicated than this simple example. A realistic
coagulation kernel will be, at the very least, a product of a mutual cross
section $\pi (r + r^\prime)^2$ and a relative velocity

\begin{equation}
\Delta V(m,m^\prime) = \sqrt{ (U - U^\prime)^2 + (V - V^\prime)^2 + 
(W - W^\prime)^2 + v_T^2(m,m^\prime)},
\end{equation}

\noindent
where $U(m)$, $V(m)$, and $W(m)$ are the $x,y,z$ systematic (pressure
gradient driven) particle 
velocities, and $v_T$ is the turbulent velocity contribution 
which in general is not separable into distinct functions of
$m$ and $m^\prime$ \citep{cuz03,orm07}. The 
systematic velocities,
which are derived for a discretized particle size distribution in the 
appendix,  are complicated functions of the particle size through the 
stopping time $t_s$, 

\begin{equation}
t_s = \frac{m\Delta V_{pg}}{F_D},
\end{equation}

\noindent
where $\Delta V_{pg} = |\vec{\bf{V}} - \vec{\bf{v}}|$ is the relative
velocity between the particle and the gas, and 
$F_D$ is the drag force on the particle of size $r$ which depends on the size
of the particle relative to the mean free path $\lambda$ of the gas molecule
\citep[e.g., see][]{cuz93}. Depending on whether $r \gtrsim
\lambda$ or $r \lesssim \lambda$ determines whether the stopping time itself 
depends on the instantaneous relative velocity $\Delta V_{pg}$
between the particle and the gas (Stokes regime) or does not (Epstein regime).
In the latter case, calculation of the systematic velocities ($U,V,W$) and
gas velocities ($u,v,w$) for a particle size distribution is
straightforward. In the former case, iterations are required to correctly
calculate $\Delta V_{pg}$ (see appendix).

The turbulent velocities are less straightforward to implement than their
systematic counterparts because of the different coupling that exists 
between particles and eddies of different sizes (and thus stopping and decay
times, respectively). It has not been until very recently that closed form
expressions for the turbulence-induced velocities (for a particle size
distribution) were derived \citep{orm07}. These expressions can be written
separately for the so-called ``class I'' and ``class II'' 
eddies\footnote{The concept of eddy classes were introduced by \citet{vol80}
to distinguish between slowly and rapidly varying eddies. Class I eddies
are defined as those in which the eddy fluctuates slowly enough that a 
particle's
stopping time $t_s$ is much shorter than the eddy decay time and the time 
it takes
to cross the eddy; thus, they will align themselves with the gas motions of 
the eddy prior to its decay.
Class II eddies then are defined as ones in which the eddy decay time is fast
compared to the particles $t_s$, and thus only provide a
small perturbation to the particle motion.} as:

\begin{equation}
\Delta V_I^2 = v_g^2\frac{{\rm{St}}_i-{\rm{St}}_j}{{\rm{St}}_i+{\rm{St}}_j}
\left(\frac{{\rm{St}}^2_i}{{\rm{St}}^*_{ij}
+{\rm{St}}_i} - \frac{{\rm{St}}^2_i}{1+{\rm{St}}_i} - \frac{{\rm{St}}^2_j}
{{\rm{St}}^*_{ij}+{\rm{St}}_j} +
\frac{{\rm{St}}^2_j}{1+{\rm{St}}_j}\right),
\end{equation}

\begin{equation}
\Delta V^2_{II} = v_g^2\left( 2({\rm{St}}^*_{ij} - {\rm{Re}}^{-1/2}) +
\frac{{\rm{St}}^2_i}{{\rm{St}}_i+{\rm{St}}^*_{ij}} - \frac{{\rm{St}}^2_i}
{{\rm{St}}_i+
{\rm{Re}}^{-1/2}} + \frac{{\rm{St}}^2_j}{{\rm{St}}_j+{\rm{St}}^*_{ij}} -
\frac{{\rm{St}}^2_j}{{\rm{St}}_j+{\rm{Re}}^{-1/2}}\right),
\end{equation}

\noindent
where $v_T^2 = \Delta V^2_I+\Delta V^2_{II}$, ${\rm{Re}}$ is the Reynolds 
number, $v_g = \alpha^{1/2}c_g$ is the turbulent gas velocity with $c_g$ the 
sound speed \citep[e.g.,][]{nak86,cuz06}, and the particle Stokes numbers 
${\rm{St}}_i = t_{{\rm{s}}i}/t_L$, where $t_L$ is the turnover time of the 
largest eddy, typically
taken to be $\Omega^{-1}$. The boundary between class I and class II is
defined by the ``combined'' Stokes number
${\rm{St}}^*_{ij} = {\rm{max}}({\rm{St}}^*_i,{\rm{St}}^*_j)$
and the values of the ${\rm{St}}^*_k$ are obtained from the
equation \citep{orm07}:

\begin{equation}
\frac{2}{3}y_k^*(y_k^*-1)^2 - \frac{1}{1+y_k^*} = -\frac{{\rm{St}}_k}
{1+{\rm{St}}_k} + \frac{1}{{\rm{St}}_k}\frac{\Delta V^2_{pg}}{v^2_g},
\end{equation}

\noindent
where $y^*_k = {\rm{St}}^*_k/{\rm{St}}_k$.
In our calculations, we will make use of these expressions when comparing
cases with turbulence-induced velocities.

Given that the moments method requires that we
express the kernel in terms of fractional or integer moments, the problem
becomes expressing the relative velocity (Eq. 10) in a form that satisfies 
this requirement. As an
example, the systematic velocities $U$, $V$, and $W$ can each be fit easily 
by a finite series in fractional powers $p_i$ of $m$
as $U(m) = \sum^N_i a_i m^{p_i}$ where the 
coefficients $a_i$ can be found by fitting $N$ points to the function 
$U(m_l) = U_l$ ($l$ an integer). This leads to the system of equations

\begin{equation}
U_l = \sum_{i=1}^N a_i m_l^{p_i}.
\end{equation}

\noindent
The coefficients $a_i$ then
follow from matrix inversion. Similarly, one can find expressions 
$V(m) = \sum_i^N b_i m^{p_i}$, and $W(m) = \sum_i^N c_i m^{p_i}$ 
so that we may
construct the laminar expression for $(\Delta V)^2$ in terms of the variables 
$m$ and
$m^\prime$ by multiplying out the individual terms, e.g., 
$U(m)U(m^\prime) = \sum_{i,j} a_ia_j m^{p_i}m^{\prime p_j}$ and so on. We
find then that we can express the square of Eq. (10) as

\begin{equation}
(\Delta V(m,m^\prime))^2 = \sum_{i,j} A_{ij} \left[m^{p_i+p_j} - 
2m^{p_i}m^{\prime p_j} + m^{\prime p_i+p_j}\right],
\end{equation}

\noindent
where $A_{ij} = a_ia_j + b_ib_j + c_ic_j$, and $p_i+p_j\le 2$ for closure. 
This determines the matrix of
coefficients $A_{ij}$ which operate on the finite power series in
$m$ and $m^\prime$. Although we cannot use the same approach for solving
for $\Delta V$ in
Eq. (10) because of the radical, we were motivated to express $\Delta V$ in 
a similar form, i.e., as 
$\Delta V(m,m^\prime) = \sum_{i,j} A^\prime_{ij} [m^{p_i+p_j} - 
2m^{p_i}m^{\prime p_j} + m^{\prime p_i+p_j}]$, and $p_i+p_j\le 1$.
The problem reduces to solving for the 
coefficients $A^\prime_{ij}$ directly, by similar matrix 
inversion techniques. Unfortunately, the number of points needed to obtain
an accurate representation of the two-variable function 
$\Delta V(m,m^\prime)$ this way 
exceeds the limitations of the inversion. In
practice we found that we could solve a system for $\Delta V$ 
with at most 
$5-6$ points ($25-36$ matrix elements),
whereas Eq. (16) carries much less restriction in the
number of points needed because we could construct $(\Delta V)^2$ 
by the product of accurate representations of single variable functions. 
Furthermore, due to the coupling evident in
the turbulence induced velocities (Equations 12 and 13), the turbulent 
velocities are not separable functions of the masses $m$ and $m^\prime$, 
further complicating the analysis.

{\it Powerlaw assumption}: Fortunately, it turns out that, although 
the direct calculation approach to $\Delta V$ described above would
be mathematically appealing in the sense that the moment equations would
remain implicit (i.e., the form of the mass distribution would only
be assumed at $t=0$), it is not necessary. In the next sections
we describe two alternative approaches to dealing with realistic coagulation 
kernels that employ the moments method under the assumption that the
form of the mass distribution $f(m,t)$ is a powerlaw.
Such an assumption is 
not entirely unfounded. A number of detailed models by 
\citet{wei97,wei00,wei04}
have shown that powerlaw size distributions result, which have 
nearly constant mass per decade radius to an upper limit $m_L(t)$ which
grows with time until the frustration limit of around a meter is reached, and
(under turbulent conditions, at least)
growth stalls \citep{cuz06}. Similar trends are found by \citet{dul05} in
which different assumptions about collisional ejecta are made. These authors
do find minor fluctuations in the distribution, but if one is primarily
interested in general properties of the distribution, such as
how the largest particle size changes with time, and not the fine 
structure of the 
mass distribution itself, the approach is quite advantageous. There are reasons
to believe that a powerlaw is a natural end-state, especially those
with equal mass per decade, because they have self-preserving properties for
the collision kernels of interest \citep{cuz06}.

The significance of the upper mass cutoff $m_L$ depends on the assumed slope 
of the powerlaw. In all real distributions, there will be a rapidly decreasing 
abundance of particles for masses exceeding some threshold, even though the 
abundance may not drop immediately to zero as in our assumed model. The rare, 
extremely large particles might be of interest for some applications,  but 
our focus will be the particle size carrying most of the area, or most of 
the mass (the first or second moments). For powerlaw distributions 
$f(m) \propto m^{-q}$ with $q < 2$, $m_L$ is itself the mass of the particle 
carrying most of the mass and is independent of the selection of a lower 
particle size cutoff. For $q=2$, the distribution contains equal mass per 
decade, and for $q>2$, most of the mass is in the small particles and the 
value of $m_L$ depends on the lower particle size chosen. Most realistic 
distributions, and those most commonly treated in the literature, have $q<2$; 
here we assume $q=11/6$, a widely used fragmentation powerlaw. In this case, 
a strict cutoff at $m_L$ represents a well-defined distribution with 
easily-understood moments where most of the mass is at $m_L$ and the area 
is nearly equally distributed per decade with a mass dependence $m^{-1/6}$. 

Although we do not include either imperfect sticking or fragmentation at this 
stage in the model, we believe that the moment equations as expressed in 
Eq. (5) will remain valid up to a "fragmentation barrier", which may be 
defined as that size for which the typical disruption energy of a particle 
is on the same order as the energy of identical colliding particles.
The fragmentation barrier will also depend on one's choice of
nebula parameters. This treatment (up to the fragmentation size) is 
consistent with recent work by \citet{bra08} (their Fig. 13) which shows
a constant powerlaw mass distribution up to a cutoff size which then falls
off abruptly. This ``knee'' in the distribution represents that efficient
fragmentation size.

Once $m_L$ reaches the fragmentation barrier, growth beyond this 
stage would need to be treated in a different manner, for
example, by simple sweepup of small, less disruptive particles by large
ones \citep{cuz93}. Creation and disruption of these
particles can be handled as part of the source and sink terms in which
for example, disrupted particles are assumed to be fragmented back into
a powerlaw distribution (which is suggested by experimental evidence and
widely assumed by other modelers) as opposed to monomers. The model as
presented within this paper, however, should be useful for the early
stages of protoplanetary nebula particle growth relevant to spectral
energy distributions and MRI suppression. We leave the incorporation of
growth stages beyond the efficient fragmentation stage for a later paper.

\subsubsection{Approach 1: Explicit Assumption}

Motivated by our discussion of the previous section, if we assume that the
form of the particle mass distribution function remains a powerlaw 
at all times, we may express $f(m,t) = c(t)m^{-q}$, such that 
$m_L(t)$ is the growing upper limit of the distribution, $q$ is the slope
which is assumed to be constant (although see below), and 
$c(t)$ is a normalization coefficient. Taking the lower limit of the mass 
distribution to be $m_0$, 
then moments as expressed in Eq. (2) are explicitly given for $q-p\neq 1$ by

\begin{equation}
M_p(t) = \frac{c(t)}{1 + p -q}\left(m_L(t)^{1+p-q} - m_0^{1+p-q}\right).
\end{equation}

\noindent
We then take the time derivative of Eq. (17) for the zeroth and second
moments, and substitute these expressions on the LHS of the corresponding 
moment equations in Eq. (5) to get

\begin{equation}
\frac{m_L^{1-q} - m_0^{1-q}}{1-q}\frac{dc}{dt} + cm_L^{-q}\frac{dm_L}{dt} =
-\frac{1}{2}c^2\Gamma_0(m_L),
\end{equation}

\begin{equation}
\frac{m_L^{3-q} - m_0^{3-q}}{3-q}\frac{dc}{dt} + cm_L^{2-q}\frac{dm_L}{dt} =
c^2\Gamma_2(m_L),
\end{equation}

\noindent
where Eq. (18) is valid for $q\neq 1$, and $\Gamma_0$ and $\Gamma_2$ are the 
integrals on the RHS of Eq. (5) for $k=0$ and $k=2$, respectively. That is

\begin{equation}
\Gamma_0(m_L) = \int_{m_0}^{m_L(t)}\int_{m_0}^{m_L(t)}
K(m,m^{\prime})m^{-q}m^{\prime -q}\,dm\,dm^\prime,
\end{equation}

\begin{equation}
\Gamma_2(m_L) = \int_{m_0}^{m_L(t)}\int_{m_0}^{m_L(t)}
K(m,m^{\prime})m^{1-q}m^{\prime 1-q}\,dm\,dm^\prime,
\end{equation}

\noindent
where the kernel is given by
$K(m,m^\prime) = \sigma(m,m^\prime)\Delta  V(m,m^\prime) S(m,m^\prime)$, with
$\sigma(m,m^\prime) = K_0(m^{1/3}+m^{\prime 1/3})^2$,
$K_0 = \pi(3/4\pi\rho_s)^{2/3}$, and $\rho_s$ is the particle material 
density, assumed
constant. Note that for the special case of $q=1$, Eq. (18) has a slightly
different form which depends on $\ln{(m_L/m_0)}$.
After some simple algebra, Eq.'s (18) and (19) can be written as

\begin{equation}
\frac{dm_L}{dt} = c\left[\frac{(3-q)(m_L^{1-q}-m_0^{1-q})\Gamma_2 +
\frac{1}{2}(1-q)(m_L^{2-q}-m_0^{2-q})\Gamma_0}{(3-q)(m_L^{1-q}-m_0^{1-q})
m_L^{2-q} - (1-q)(m_L^{3-q}-m_0^{3-q})m_L^{-q}}\right],
\end{equation}

\begin{equation}
\frac{dc}{dt} = -c^2\left[\frac{(3-q)(1-q)(m_L^{-q}\Gamma_2 +
\frac{1}{2}m_L^{2-q}\Gamma_0)}{(3-q)(m_L^{1-q}-m_0^{1-q})
m_L^{2-q} - (1-q)(m_L^{3-q}-m_0^{3-q})m_L^{-q}}\right],
\end{equation}

\noindent
which we integrate using a fourth order Runge-Kutta scheme. Equation (17) then
gives the moments $M_0$ and $M_2$ as a function of time, which can be
directly compared with direct integration of the same conditions
using the coagulation equation (Eq. 1). 

The advantage of this approach (in which a powerlaw is assumed at all
times) is its transparency; that is, the variables being sought
($m_L$ and $c$) are solved
for directly. Furthermore, the change in the coagulation kernel as the
particle size distribution changes is included because the kernel is
updated and explicitly integrated into $\Gamma_0$ and $\Gamma_2$ with every 
time step. This will prove
advantageous when additional effects such as sticking
are included in the kernel. In addition, source and sink terms need not
parameterized in terms of integer moments and can be implemented directly.
An unfortunate disadvantage of this approach is
that because the kernel must be integrated (in fact several times) over
both $m$ and $m^\prime$ every
time step to get $\Gamma_0$ and $\Gamma_2$, the CPU time involved is 
significantly longer than a fully
implicit case (e.g., {\S} 2.1; also see {\S} 2.2.2); however, it
remains a much faster approach (orders of magnitude) than solving Eq. (1) 
directly since the
cumbersome convolution has been eliminated.

\subsubsection{Approach 2: Semi-implicit Assumption}

Here, we present an alternative moment-based approach to solving the realistic
coagulation kernel in which the integer moments appear directly. 
Unlike the previous case of {\S} 2.2.1, where the double integral of
Eq. (5) is used to get the functions of $m_L$, $\Gamma_0$ and $\Gamma_2$, 
we only integrate over one mass
variable (i.e., only integrate one of the integrals), defining the
functions

\begin{equation}
C_0(m,t) = \int_{m_0}^{m_L} K(m,m^\prime)f(m^\prime,t) \,\,
dm^\prime,
\end{equation}

\begin{equation}
C_2(m,t) = \int_{m_0}^{m_L} m^\prime K(m,m^\prime)f(m^\prime,t) 
\,\,dm^\prime,
\end{equation}

\noindent
where the form of the kernel $K(m,m^\prime)$ is the same that in {\S} 2.2.1.
We then fit $C_0$ and $C_2$ with a finite series in fractional
powers of $m$ in the same manner as given in Eq. (15). Substituting these
functions in place of one of the integrals (say over $m^\prime$), we may
integrate over $m$ to get

\begin{eqnarray}
\frac{dM^\prime_0}{dt} = -\frac{1}{2M_0(0)}\int_{m_0}^{m_L(t)}
f(m,t)C_0(m,t)\,dm = -\frac{1}{2}\sum_i a_i \mu_{p_i}M^\prime_{p_i} \nonumber \\
\frac{dM^\prime_2}{dt} = \frac{1}{M_2(0)}\int_{m_0}^{m_L(t)} m f(m,t)C_2(m,t)
\,dm = \sum_i b_i \nu_{p_i+1}M^\prime_{p_i+1}, \nonumber \\
M^\prime_{p_i} = \left[M^\prime_0\right]^{\frac{1}{2}(p_i-1)
(p_i-2)}\left[M^\prime_1\right]^{-p_i(p_i-2)}\left[M^\prime_2\right]^
{\frac{1}{2}p_i(p_i-1)},
\end{eqnarray}

\noindent
where we have made use of equations (8) and (9) to express the solution in
terms of the integer moments $k = 0,1,2$. Here, $\mu_{p_i} = M_{p_i}(0)/
M_0(0)$, $\nu_{p_i+1} = M_{p_i+1}(0)/M_2(0)$, $M^\prime_k = M_k(t)/M_k(0)$,
and $0\leq p_i \leq 1$. The $C_k$ are fairly smooth functions over a large
range of particle radii;, however, the accuracy of fitting a single series in 
fractional moments over a very broad range of particle sizes 
(i.e., over many orders of magnitude) may drop off significantly as the 
broadness of the range increases. This issue may be circumvented by 
employing a piecewise fit to the integrated kernels $C_k$.

It is interesting to note that by this definition of $C_k$, we have
effectively accomplished what we set out to do in
our discussion at the beginning of {\S} 2.2, that is, defining the 
coagulation kernel in terms of finite series in powers of the mass $m$. 
The difference
here is that we have done so through the first integral of the kernel, and 
not the kernel itself. This means that the mass distribution $f(m,t)$ is
expressed in the calulation of the $C_k$, but remains
implicit in the definition of the ODEs (Eq. 26). Thus, the method is
semi-implicit, because the RHS of the equations above can be
expressed in terms of the moments (as defined in Eq. 2). Equations (26) 
are then integrated using the fourth order Runge-Kutta method, and may be
compared with the results of {\S} 2.2.1 and the direct integration of
Eq. (1).

This semi-implicit approach tracks
the evolving kernel through the integration of Equations (24) and (25)
after every timestep, thus the computational time involved is similar
to the explicit case. In order to update the kernel,
one may solve for the new $m_L$ after each $\Delta t$ 
using the equation ($q\ne 1$)

\begin{equation}
\frac{m_L-m_0}{m_L^{2-q}-m_0^{2-q}} - \frac{M_q}{(2-q)M_1} \simeq 0.
\end{equation}

\noindent
The new
normalization coefficient $c$ can then be found from the definition of 
$M_1 = \rho$.
We then reintegrate Equations (24) and (25) under the powerlaw assumption, 
and then proceed to fit the $C_k(m,t)$ with a finite series in fractional 
powers of $m$. Although, in principle, any other two moments could be
used to obtain $m_L$, $M_q$, which lies between $M_2$ and $M_1$, and because
it roughly characterizes the evolution of the largest particle (see discussion
at the end of {\S} 2.2), seems the most
consistent choice. The $q$-th moment is calculated using the Lagrange 
polynomial interpolation scheme (Eqs. 8 and 9).

The advantage of the moments method lies in the ability to express the
differential equations in terms of the moments of the 
distribution (i.e., their integrated properties). 
If a more explicitly
mass-dependent approach is adopted (as is the case in {\S} 2.2.1, and the
semi-implicit approach described here), 
then the computational time significantly increases.
One can improve the speed of computation by calculating the $C_k$
periodically, or in the extreme case, only at $t=0$ which would make
the approach truly {\it implicit} (e.g., {\S} 2.1). The advantage of an 
implicit approach
is that it becomes fully general
(the form of $f$ is only assumed at the onset), and
also in the time it takes to solve ($< 1$ minute). The bulk of the time is 
spent in the
integration of equations (24) and (25) which would occur only once.
The disadvantage, of course, is that the particle velocity distribution is
not updated as it changes with time (due to, e.g., changes in the bounds
of the size distribution). We present examples of both extremes in {\S} 3.

If one wanted to implement a mass- or velocity-dependent sticking
coefficient $S$, it can readily be included in the integration to
obtain the $C_k$. The additional inclusion of source and sink terms 
due to erosion, fragmention, or
gravitational growth in this semi-implicit formalism would require that we fit 
these terms in a similar manner to the $C_k$ so that their subsequent
integration over all $m$ will yield sums over integer moments weighted by
different sets of coefficients. Caution must be exercised in fitting, e.g.,
the gravitational growth term to ensure
that the system of equations remains closed. Under such circumstances, a
fully explicit approach such as that of {\S} 2.2.1 may be preferred. 
Alternatively, these effects may be included in particle-histogram space
in between coagulation interations. 

Finally,
we should point out that allowing for other parameters (such as the index
of the powerlaw $q$)
to vary with time, does not pose a problem in either of 
the approaches we have presented. In both cases, one would simply need an
additional moment, e.g. $M_3$, to determine $q(t)$. Similarly, a bifurcated
distribution in which the powerlaw exponent changes at a particular particle
size \citep[see, e.g.,][]{ken99} may also be studied. 

In {\S} 3, we will compare the
two approaches to the direct integration of the coagulation equation for 
cases in which there are
only systematic velocities ($v_T = 0$), as well as cases in which the
velocity differences are induced by turbulent motions.

\section{Numerical Results}

We carried out several calculations in order to demonstrate the accuracy
of each alternative method compared to the brute force integration of the
collisional coagulation equation. For the purposes of comparison, 
unless otherwise noted, we chose the initial conditions to be a minimum mass 
solar nebula at 1 AU at a height of $z=10^3$ km above the midplane, and a 
particle size distribution with a minimum initial radius of 1 cm and 
a maximum initial radius of 10 cm. The
powerlaw exponent $q = 11/6$, which is assumed to be constant in these
calculations, is representative of a fragmentation population.
Standard integrations were carried out with a timestep of $\Delta t = 10$ 
years.

\subsection{Laminar Case ($v_T = 0$)}

In Figure 2, we compare the case in which there are only systematic 
(pressure-gradient driven) velocities
between particles ($v_T=0$). The solid curves represent the explicit
assumption (invariant powerlaw slope $q$ assumed in integration of 
Equations 22 and 23), while the dashed curves represent the
implicit assumption. That is, in the integration of Eq. 26, the form of
the mass distribution $f$ is assumed only at $t=0$. As before, the symbols 
represent the brute force calculation for two different grid resolutions 
(solid = 100 bins, open = 1000 bins). Since with the explicit assumption we do
not solve for the moments specifically, the values for the solid curves were
obtained by substituting the time integrated values of $c(t)$ and $m_L(t)$
back into Eq. (17) for $k = 0,2$ only. This is because
although we have used the moment equations to obtain these results, the
explicit assumption of an invariant powerlaw size distribution means that 
only the equations for the moments $M_0, M_2$ are needed (although see
{\S} 3.2). This is not true, however, for the implicit (and semi-implicit)
assumption, where 
all the moments $M_0,M_1,M_2$ appear in the differential equations.

We see that there is excellent agreement between both approaches and the
numerical values obtained with the highest resolution case. In fact, the
agreement between the explicit and implicit assumptions is also
quite good. However, at this stage there has not been a great deal of 
growth (the largest particle size in the distribution has only grown 
to $r_L = 11$ cm in size by the end of the simulation for the explicit case).
Note that even though
in the implicit case the $C_k$ are calculated only once, the estimate for the
largest mass $m_L$ for the evolving distribution $f(m,t)$ is still calculated
using Eq. (27). At least in the case of minimal or slow growth, it appears
that an implicit approach, or even periodic calculation of the $C_k$, may be
sufficient.

Numerical glitches in the low resolution brute force case arise from the 
interpolation
scheme for sampling the kernel. In particular, these glicthes are 
likely enhanced because the systematic 
relative velocities quickly approach zero
for identical particle sizes (with the effect much more prominent for larger
particles, hence not appearing so much in $M_0$). The higher resolution case
contains enough points to smooth out this effect. 

Note that (in all simulations) the direct integration of the coagulation 
equation gives a constant $M_1$
as is to be expected given that we found $dM_1/dt = 0$ in the
derivation of the moment equations in the absence of sources and sinks; this 
further validates the numerics 
of the brute force solution even
for the complicated collisional kernel being utilized, and also provides
a posteriori validation of our result that $M_1=\rm{constant}$ from,
e.g., Eq. (5), which further validates derivation of equations (24)
and (25) which was based on symmetry of the kernel.

\subsection{Turbulence Case ($v_T \neq 0$)}

Next, we explored cases in which the systematic velocities were set to zero
so that velocity differences between particles are due only to those induced
by turbulence. Depending on the magnitude of the turbulence parameter
$\alpha$, these induced velocities can be either large or small relative to the
systematic velocities. To demonstrate, we ran cases for three
different values $\alpha = 10^{-6},10^{-5}$, and $10^{-4}$. In the absence
of any mechanism to counter growth (e.g., fragmentation), larger $\alpha$
translates to faster rates of growth (due to larger relative 
velocities).

In Figure 3 we plot the results for $\alpha = 10^{-4}$,
for the explicit (solid curves) and implicit (short dashed curves) 
approaches. The growth rate is more rapid than in the laminar case, with 
the second scaled moment $\sim 70$\% larger (compared to Fig. 2) at the
end of the run, meaning that the
largest particle size achieved is $r_L \sim 13.5$ cm. We note
that, as was the case for the $M^\prime_2$ curves in Fig. 2, the explicit
and implicit cases are very similar; however, this is not the case
for the $M^\prime_0$ curves. The explicit approach overestimates the the
value of $M^\prime_0$ compared to the highest resolution brute force 
calculation, whereas the implicit approach significantly underestimates
the zeroth moment. Both approaches understimate the value of $M^\prime_2$.
When we compare integrations of $M^\prime_2$ for smaller
values of $\alpha$ as we do in Figure 4, we find that both approaches fit
the coagulation calculation quite well. The growth rate for these two
values of $\alpha$ are more in line with growth rate found when there are
only systematic velocities, so the agreement should not be surprising.

The long-dashed curves in Figures 3 and 4 
are the implementation of the semi-implicit approach ({\S} 2.2.2) in which
the $C_k$ are updated at every time step. In this case, the semi-implicit
approach provides a much a much better fit to the brute force calculation,
indicating that the kernel is evolving fast enough that an implicit
approach cannot capture this effect.
We consider the slight discrepencies between the 
semi-implicit approach and the brute force calculation to be as much a 
result of grid resolution as inaccuracy in using the Lagrange polynomial
fits to the fractional moments. The explicit and fully implicit approach 
values of $r_L$ apparently understimate the largest particle size
relative to the semi-implicit approach (which gives a value of 
$r_L \simeq 14$ cm).

Finally, we explored a variation of the explicit approach in which
the condition $M_1 = \rho$ was strictly enforced (recall that
$M_1$ does not appear in Equations 18-19). This amounts to replacing
Eq. (18) (derived for $M_0$) with the corresponding equation for $M_1$.
As a result, we cannot simultaneously fit both $M_2$ and $M_0$. 
Alternatively, we can use only Eq. (19) for the second moment, by substituting
$c = (2-q)\rho/(m_L^{2-q}-m_0^{2-q})$ (from the definition of $M_1$, Eq. 17)
in Eq. (19) so that Eq. (19) alone
determines the growth of the largest particle. The differential
equation for $m_L$ then becomes

\begin{equation}
\frac{dm_L}{dt} = \left[\frac{(3-q)(2-q)\rho\Gamma_2}
{(3-q)(m_L^{2-q}-m_0^{2-q})m_L^{2-q} - (2-q)(m_L^{3-q}-m_0^{3-q})
m_L^{1-q}}\right].
\end{equation}

\noindent
The results of the integration of Eq. (28) are shown as the dotted curves
in Figure 3 and for the $\alpha = 10^{-4}$ case in Figure 4. It is clear that 
this modification provides a much better fit to $M_2$, perhaps even better
than the semi-implicit case above. However, using
$m_L(t)$ calculated from Eq. (28), and solving for $c(t)$ does a poor job 
fitting $M_0$ (see Figure 3). Thus we conclude that although the
modification of the explicit approach matches the brute force calculation of
$M_2$ quite well, only the semi-implicit case is able to provide a
simultaneous fit to both the zeroth and second moments (under the
condition that $M_1=\rho = {\rm{constant}}$).

\subsection{A Model Comparison}

\citet{gar07} has developed a simplified analytical approach for dealing 
with the
growth of particles in a turbulent regime, in which the particle size
distribution is parameterized by a powerlaw in particle mass with the same 
exponent we have used in this paper ($q = 11/6$).
Similar to what we have presented in previous sections, the underlying
assumption of \citet{gar07} is that collisions between particles occur
frequently enough that a steady-state balance is reached in the form of
the particle
size distribution, but with an upper size cutoff that varies with time.
Thus, the model of \citet{gar07} is {\it explicit} and follows only two 
parameters: the growth
of the largest particle $m_L(t)$, and the normalization factor $c(m_L,t)$.

Given the similarity of the underlying assumptions for the Garaud model
and the examples we have presented, it is a
useful exercise to compare the results of the two approaches directly.
The growth of the largest particle $r_L$ in the Garaud model (her Eq. 36), 
expressed in terms of the notation used in this paper, is given by

\begin{equation}
\frac{dr_L}{dt} = \frac{\rho \Omega H}{\rho_s} \sqrt{\frac{\alpha
\gamma {\rm{St}}_L}{1 + 64{\rm{St}}^2_L(2 + 5{\rm{St}}_L^{-0.1})^{-2}}}.
\end{equation}

\noindent
In Eq. (29) ${\rm{St}}_L$ is the Stokes number for the largest particle 
$r_L = (3m_L/4\pi\rho_s)^{1/3}$, $H=c_g/\Omega$ is the scale height of the gas,
$\gamma$ is the adiabatic index of the gas, and
in this expression it is assumed that the sticking coefficient $S = 1$, and 
that $m_L>>m_0$. The above
equation is similar to the formula for grain growth proposed by \citet{ste97}
to factors of order unity. Finally, we point out that 
the Garaud model for particle growth
is restricted to the value $q = 11/6$ in order to preserve its completely
analytical nature. 
As a means of a fair comparison, we chose to compare the Garaud model to
our explicit case (Eq. 28) for reasons explained below. In the limit of
$m_L >> m_0$ and $q=11/6$, Eq. (28) becomes

\begin{equation}
\frac{dr_L}{dt} \simeq 0.05 \frac{\rho}{\rho_s}\frac{\Gamma_2}{m_L}.
\end{equation}

We present
the results of our comparison in Figure 5 for the same initial conditions
as described at the beginning of {\S} 3.2 (upper curves). 
We find quite generally that our explicit calculation (Eqs. 28 and 30)
leads to a faster growth rate than what is predicted from the Garaud model,
initially. We note that
the Garaud expression steepens quickly for $r_L \gtrsim 13$ cm,
suggesting that the growth rates of the two approaches are more comparable at 
later times. 
However, the minor ``kink'' in
the long-dashed curve is due to the shift from Epstein to Stokes flow. A
much more subdued kink is visible in the explicit approach
which uses the full expressions for the turbulent velocity, whereas in the
derivation of Eq. (29), it is
assumed that the stopping time in the Stokes regime is defined
by some mean characteristic velocity which leads to a much more
noticeable discontinuity.
Regardless, the overall more subdued growth rate elicited by Eq. (29) (despite 
the steepening at later times due to a shift in flow regimes) is apparent 
from the lower set of curves in Fig. 5 where
the initial conditions were chosen with $r_L(0) = 1$ cm. For this case, growth
occurs only in the Epstein regime, but still, the curves for the explicit 
approach
and that calculated from Eq. (29) begin to diverge. Thus,
the Garaud expression apparently underestimates the growth rate relative to
our approach.

We emphasize that the treatment of particle growth by \citet{gar07} requires
several approximations in order to derive a purely analytical expressions for
$dr_L/dt$. Besides the aforementioned restriction of $q=11/6$, Garaud 
approximates the full expressions for the turbulent velocities that we
use here by partitioning $v_T$ into seperate cases dependent on
the particles' stopping time relative to the turnover times of the smallest
and largest scale eddies. Furthermore, some question may be raised as to the
comparability of the moment equations used here to derive Eq. (28), versus
the particle growth equation used by \citet[her Eq. 29]{gar07}. 
The equation used by \citet{gar07}
is more akin to a ``sweep-up'' equation, with no sources or sinks, than to a
formal coagulation equation. Because
it bears some resemblance to the equation for $M_2$ (with some algebra,
to factors of order unity), we concluded that our Eq. (28) is the appropriate 
analog. Despite the differences in growth rate, we find the agreement in
the general trend of growth of the two approaches reassuring.

\section{Opacity Calculations}

Evolutionary models of protoplanetary nebulae, giant planet atmospheres, etc.
must somehow treat the escape of thermal radiation \citep{pol96,hub05,dur07}.
In most cases, particles provide the primary opacity for these
models. Observations of these and similar objects often rely on Spectral Energy
Distributions (SEDs) which can be compared to a model once the model's internal
temperature distribution is known; clear evidence is seen for grain growth in
many cases \citep[see review by][]{nat07}. Because of the nearly 
insurmountable
computational burden involved with performing a fully self-consistent
calculation of particle growth by coagulation along with an already difficult
fluid dynamical calculation, most modelers simply assume some invariant
particle size distribution, such as the MRN interstellar grain distribution, or
make arbitrary assumptions about particle growth \citep{hub05}.

In the simplest regime (monodisperse particle radius $r$ larger than a
wavelength), the particle opacity can be written as the area per unit 
mass:

\begin{equation}
\kappa = \frac{3}{4 \rho_s r} \hspace{0.1 in} {\rm cm}^2\,{\rm g}^{-1};
\end{equation}

\noindent
thus growth in radius from 0.1$\mu$ to 1 mm leads to a factor of $10^4$ change 
in opacity. To the degree that this wavelength-independent regime holds,
including particle size evolution by the moments method in one's
evolutionary models would allow a very simple way to track particle growth and
decreasing opacity. For instance, equation (31) above is easily generalized to
the area per unit mass integrated over the size distribution:

\begin{equation}                       
\kappa = \frac{\int \pi r^2 f(m)\, dm}{\int m f(m)\, dm} 
       = \left(\frac{9 \pi}{16 \rho_s^2}\right)^{1/3}\frac{M_{2/3}}{M_1}.
\end{equation}

\noindent
As an application of the moments method in Figure 6, we have calculated 
the decrease
in opacity (given by Eq. 32) with time using the semi-implicit approach.
An initial particle size distribution with a lower bound of $r_0=0.1$ cm
and $q=11/6$ was used. Both the pressure gradient driven systematic 
velocities and the
turbulence-induced velocities were used. In the absence of any mechanism to
hinder particle growth, larger values of $\alpha$ lead to more steeply
decreasing opacities with time.

In a regime where the particle extinction efficiency $Q(r,\lambda)$ is
wavelength-dependent (say, if the particles are comparable to or smaller than
the wavelength), one simply integrates $Q(r,\lambda)$ over the powerlaw mass
distributions resulting from the moments model.  For example,
\begin{eqnarray}
\kappa_{\lambda} = \frac{\int_{m_0}^{m_L} \pi r^2 Q(r,\lambda) f(m)\, dm}
             {\int_0^{m_L} m f(m)\, dm}
   = \frac{1}{M_1}\int_{m_0}^{m_L} \pi r^2 Q(r,\lambda) c(m_L)m^{-q}\, dm 
\nonumber \\
   = \frac{c(m_L)}{M_1}\left(\frac{9 \pi}{16 \rho_s^2}\right)^{1/3}
\int_{m_0}^{m_L} Q(r,\lambda)m^{2/3 -q } dm. 
\end{eqnarray}

\noindent
These opacities $\kappa_{\lambda}$ can be used to calculate Planck or Rosseland
(wavelength-averaged) means for use in radiative transfer models.
Recall that the powerlaw slope
$q$ can be freely adjusted within a small but plausible range to explore
different growth regimes.

\section{Porosity}

Fractal growth of particles by low-velocity sticking of small solid monomers
with radius $r_o$ and mass $m_o$ (and/or aggregates of such monomers) causes
them to have a density much less than the material density of the monomers
\citep{bec00,dom07,orm08}. These porous
particles can be described as fractals with dimension $D$, such that the
particle mass $m$ increases proportionally to $r^D$ where $r$ is some effective
radius and $D$ is the fractal dimension. Thus the particle's internal density
is a function of particle size: 

\begin{equation}
\rho(r) = \frac{3 m }{4 \pi r^3} \sim \frac{3 m_o (r/r_o)^D}
{4 \pi r^3} = \frac{3 m_o}{4 \pi r_o^3}(r/r_o)^{D-3} = \frac{3 \rho_o}
{4 \pi}(r/r_o)^{D-3}.
\end{equation}

\noindent
For a typical situation where $D \sim 2$ \citep{dom07}, 
$\rho(r) \propto r^{-1}$ and thus the
product $r \rho(r)$ is a constant across a wide range of particle sizes (until
compaction sets in). These more complex but quite plausible particle
density-size relationships complicate the expressions for particle stopping
time and Stokes number ({\S} 2.2, Eq. 11; also, see appendix). 

Because the particle stopping times enter in through the collisional kernel, 
the fractal nature of particles can be accounted for using the method of 
moments in a straightforward manner while maintaining the criterion for 
closure of the system. We can verify this by noting that the mutual particle 
cross section 

\begin{equation}
\sigma (m,m^\prime) = \frac{\pi r_o^2}{m_o^{2/D}}\left(m^{1/D} +
m^{\prime 1/D}\right)^2,
\end{equation}

\noindent
is proportional to $m^{2/D}$, whereas the stopping time (Eq. 11) in the Epstein
regime (the regime that would apply to fluffy fractal aggregates) where
the drag force $F_D = (4/3)\pi r^2 c_g \rho_g \Delta V_{pg}$ 
\citep[e.g.,][]{cuz93} is given by 

\begin{equation}
t_s = \frac{3m}{4\pi r^2 c_g\rho_g} = \frac{3m_o^{2/D}}{4\pi r_o^2 c_g\rho_g}
m^{1-2/D}.
\end{equation}

\noindent
In the limit of small Stokes number
($\rm{St}\ll 1$), both the systematic and turbulence-induced velocities
are proportional to $\rm{St}$, so that $\Delta V \propto t_s \propto 
m^{1-2/D}$, and the entire kernel is proportional to $m$. For 
larger $\rm{St}$, the kernel has a shallower powerlaw
dependence. Thus, in general, the dependence of $K(m,m^\prime)$ on the
mass is $\lesssim m$, preserving closure of the system, even for fractal
particles. Even though the variation in the properties of the evolving 
distribution due to porous particles are incorporated directly into the 
kernel, and are folded into the integration of the explicit approach, the
effects of fractal aggregates can, nonetheless, affect 
which moments characterize what properties in the semi-implicit (or implicit)
approach. For
example, the wavelength-independent opacity expressed in Eq. (32) takes
the form $\kappa = (\pi r_o^2/m_o^{2/D})M_{2/D}/M_1$.

It should be noted that the value $D=2$ represents a special case in that
the particle-to-particle relative velocities in the Epstein regime do not
depend on the mass of the fractal particle. Indeed, this would seem to
indicate that fractal growth can proceed unabated with the corresponding
stopping time of the fluffy aggregate remaining the same as that of a single 
monomer, which would have a significant effect on other particle properties. In
particular, the wavelength-independent opacity for $D=2$ is constant. However,
impacts will eventually lead to compaction or even fragmentation depending
on the relative velocities \citep[e.g.,][]{orm07b}. 
Both fractal grains and non-fractal particles in the same mass distribution 
can be treated in the explicit approach without any modifications, while a 
piecewise fit to 
the integrated kernel $C_k(m)$ ({\S} 2.2.2) can be used in order to 
account for the change in regimes in the semi-implicit approach.

\section{Conclusions}

We have demonstrated an approach to solving the collisional coagulation
equation with an arbitrary collisional kernel which should be useful in cases
when it is only necessary to keep track of general
properties of the distribution. This approach involves solving a finite set of
coupled differential equations in terms of the integer moments of the particle
size distribution. The number of equations (and thus moments) needed depends
on the number of properties being tracked. The advantage of the moments method 
approach is that it allows for considerable savings in computational
time compared to direct integration of the coagulation equation,
which requires keeping track of every particle size at every spatial location
and timestep. 

In this paper we have specifically
studied the growth of the largest particle under the assumption that
the particle size distribution is a powerlaw; however, the technique can be
extended to track other properties of the distribution that may change with
time. There are many reasons to believe that a powerlaw size distribution is
a natural end-state of particle growth, especially those with equal mass
per decade, because they have self-preserving properties \citep{cuz06}.
With the assumption of a powerlaw distribution, we
have provided two different approaches to solving the moment equations,
one explicit in which the powerlaw assumption is enforced rigorously at
all times, and a semi-implicit approach in which the kernel is integrated
over one of the mass variables as much as once every time step. The latter 
approach can be made fully implicit by only assuming the form of the mass
distribution at $t=0$. These approaches
are significantly faster than solutions of the coagulation equation
because, in particular, the convolution integral (first term on RHS of Eq. 1) 
has been eliminated. In realistic evolutionary models, intermediate steps
performed in particle ``histogram'' or ``size-distribution'' space may be
interleaved with moments-based coagulation steps in order to account for,
e.g., advective/transport terms.

We have compared
these alternate approaches to the brute force integration of the full
coagulation equation for cases in
which there are only systematic velocities, and cases in which the differences
in velocity between particles is induced by turbulence. If the growth rate
is gradual, the explicit and implicit
approaches match the brute force calculation well (Fig. 2), 
whereas
faster growth rates are more difficult to model (Fig. 3). We find that we 
are able to use the semi-implicit approach in which the $C_k$ are updated
at every timestep, and a modification to the explicit approach, in which
we solve the equation for $M_2$ only with the assumption 
of $\rho=M_1$ strictly enforced, 
in order to compensate for the faster growth rates.
The modification to the explicit approach is useful if
one is not particularly interested in following the evolution of the number
density of particles, or other properties which may be decribed by 
(fractional) moments $< 1$ (e.g., see {\S} 4).  Our results also suggest
that a fully implicit approach is probably most useful under circumstances in
which the kernel depends on the mass in a straightforward manner (e.g, the
Saffman-Turner kernel, {\S} 2.1).

We have compared the approaches developed in this paper to an alternative 
model for particle growth in a turbulent nebula
\citep{gar07}. We have found that there is fairly good agreement in general,
but that the curves diverge as time proceeds. This does not appear to be
particle-size dependent, or due to a shift in the flow regime.
The Garaud expression (Eq. 29) underestimates the growth rate of 
the largest particle size relative to our approach by only $\sim 20-30$\% 
in our comparison. We note that the advantage 
of the method of \citet{gar07} is 
that it is purely analytical; however, preservation of her analytical
approach requires, amongst other things, the powerlaw 
exponent be restricted to $q=11/6$. Our approach has no such restriction
on the choice of exponent $q$, nor for $q$ to even be a constant.

As a sample application, we show how the moments method can
be used in one's evolutionary model to track
particle growth and opacity. As a specific case, we calculated
the change in wavelength-independent opacity with time for an initial
particle size distribution with upper and lower bounds of $0.1-1$ cm. 
Both systematic and turbulent velocities were included. In the absence of
any mechanism to counter growth, the opacity decreases sharply for higher
choices of the turbulent parameter $\alpha$. Extension to cases in
which the extinction efficiency is wavelength dependent is straightforward.
Such opacities can be used to calculate Planck or Rosseland 
wavelength-averaged means for use in radiative transfer codes.

Finally, we indicate how porous particles with fractal dimension $D$ can
be accounted for in the moments method. The particle-size density
relationships that arise affect the particle stopping times and Stokes
number, both of which appear only in the collisional kernel. Thus
implementation is straightforward. We can treat mass distributions composed
of only fractal grains, or both fractal and non-fractal particles.
In either case, relatively little modification is needed in the explicit
approach, whereas with the semi-implicit (and implicit) approach, a piecewise
fit to the integrated kernel $C_k$ using the method as described in {\S} 2.2.2
can be used to account for the change in particle growth regime when both
types of particles are included.

The computational burden of directly solving the coagulation equation
makes it
quite prohibitive to explore large regions of parameter space, and thus 
it serves
as the primary bottleneck in evolutionary growth models. The approach 
demonstrated herein is intended to obtain robust, quantitative results for 
disk properties such as particle growth 
timescales and ``typical'' particle sizes that may be used in modeling 
efforts that are
focused more on the larger problem of planetesimal formation.

Although we have not included sticking or fragmentation in this paper, we
believe that the moment equations (Eq. 5) will remain valid at least up
to the fragmentation barrier. This size will depend on the
assumed particle strengths and choice of nebula parameters. The treatment
of growth up to the fragmentation size is consistent with recent work by
\citet{bra08} for the case in which additional effects such as radial
drift are not included. The model as presented in this paper, however,
should be useful for the early stages of protoplanetary nebula particle
growth relevant to spectral energy distributions and MRI suppression.

In a forthcoming paper we will explore the effects of a
variable sticking coefficient $S(m,m^\prime)$. Furthermore, we will explore the
addition of source and sink terms such as gravitational growth,
erosion, sublimation, and condensation in addition to fragmentation. The 
ultimate goal will be to apply this methodology to 
a global model that studies the evolution of both the gas and solids in
nebular and subnebular (giant planetary) environments.

\acknowledgements{We wish to thank Sandy Davis, Fred Ciesla, and Olenka 
Hubickyj for internal reviews which improved the exposition of this paper,
and an anonymous reviewer for his or her careful analysis of the manuscript.
This work was supported by a grant from NASA's
Origins of Planetary Systems Program.}

\appendix

\section{Generalization of the Systematic (Pressure Gradient
  Driven) Velocities}

In this appendix, we generalize the basic equations of \citet[also, see
Tanaka et al. 2005]{nak86} for a two-component fluid by extending 
the particle component to incorporate a size distribution. 
We adopt cylindrical coordinates $(R,\phi,Z)$, designating the corresponding 
gas velocity components as
$(u,v,w)$ and the particle velocity components as $(U_i,V_i,W_i)$.
The equations of motion of
particles and gas for the radial and tangential velocities, generalized for
a particle size distribution, are given by

$$
\frac{\partial U_i}{\partial t} = -A_i \rho_g(U_i - u) + 2\Omega V_i,
\eqno{(\rm{A}1)}
$$
$$
\frac{\partial V_i}{\partial t} = -A_i \rho_g(V_i - v) - \frac{1}{2}\Omega 
U_i, 
\eqno{(\rm{A}2)}
$$
$$
\frac{\partial u}{\partial t} = -\sum_j A_j \rho_{j}(u - U_j) + 2\Omega v -
\frac{1}{\rho_g}\frac{\partial p_g}{\partial R},
\eqno{(\rm{A}3)}
$$
$$
\frac{\partial v}{\partial t} = -\sum_j A_j \rho_{j}(v - V_j) - \frac{1}{2}
\Omega u,
\eqno{(\rm{A}4)}
$$

\noindent
where the system is assumed to be axysimmetric. Here, $A_i =
(\rho_gt_{{\rm{s}}i})^{-1}$ with $t_{{\rm{s}}i}$ the stopping time of a 
particle of radius
$r_{i}$, $\rho_g$ and $p_g$ are the gas mass density and pressure, 
and $\rho_{j}$
is the material density of particles of radius $r_{j}$. We have not included 
the
vertical component equations; here, generalization to a particle size 
distribution
is straightforward since it remains true in general that $|w/W_i| << 1$; that 
is, the vertical gas component of the velocity $w$ is negligibly small
compared with $W_i$ \citep[see appendix,][]{nak86}.

If the dust stopping time is short compared to the orbital period, we can seek
steady-state solutions for the velocity components. Setting $\partial/\partial
t = 0$, this system can be solved exactly in the Epstein regime. By solving 
Eq. (A1) for $V_i$ and inserting into Eq. (A2), we obtain

$$
U_i = \frac{A_i^2\rho_g^2u + 2\Omega A_i\rho_gv}{A_i^2\rho_g^2 + \Omega^2} = 
\frac{u + 2v{\rm{St}}_i}{1 + {\rm{St}}_i^2},
\eqno{(\rm{A}5)}
$$

$$
V_i = \frac{-(\Omega/2)A_i\rho_gu + A_i^2\rho_g^2v}{A_i^2\rho_g^2 + \Omega^2}
= \frac{v - (1/2)u{\rm{St}}_i}{1 + {\rm{St}}_i^2},
\eqno{(\rm{A}6)}
$$

\noindent
where $V_i$ was obtained by insertion of $U_i$ back into Eq. (A1). In
Eqs. (A5)-(A6) we have expressed the last equality in terms of the Stokes
number ${\rm{St}}_i = t_{si}\Omega$, with $\Omega$ being the local Kepler
frequency. These expressions for the
individual particle velocity components can be inserted into Eqs. (A3) and
(A4) to yield expressions for radial and tangential gas velocity components.
With little difficulty, one finds

$$
u = 2\eta v_K \frac{s_1}{s_1^2 + (1 + s_0)^2},
\eqno{(\rm{A}7)}
$$

$$
v = -\eta v_K \frac{1 + s_0}{s_1^2 + (1 + s_0)^2},
\eqno{(\rm{A}8)}
$$

\noindent 
where

$$
s_0 = \sum_j \frac{A_j^2\rho_g\rho_{j}}{A_j^2\rho_g^2 + \Omega^2} =
\sum_j \frac{\rho_{j}}{\rho_g}\frac{1}{1 + {\rm{St}}_j^2},
\eqno{(\rm{A}9)}
$$

$$
s_1 = \Omega \sum_j \frac{A_j\rho_{j}}{A_j^2\rho_g^2 + \Omega^2} =
\sum_j \frac{\rho_{j}}{\rho_g}\frac{{\rm{St}}_j}{1 + {\rm{St}}_j^2}.
\eqno{(\rm{A}10)}
$$

\noindent
and $\eta = -(1/2\rho_g\Omega^2R)(\partial p_g/\partial R)$. These expressions
agree with those obtained by \citet{tan05} with the exception that
their expressions for $s_0$ and $s_1$ are expressed in integral form.

From these expressions, one can easily obtain expressions for the relative
velocities with respect to the gas $U_i-u$ and $V_i-v$, or the relative
velocities between particles of different sizes $U_i-U_j$ and $V_i-V_j$.
These expressions are also applicable for the Stokes regime in which the
stopping times are a function of the relative particle-gas velocities
(cf. Eq. 11). In
this case, iterations must be performed in order to obtain the correct
relative velocities. Convergence is generally achieved in a small number 
of iterations.

\clearpage

\begin{figure}
\plotone{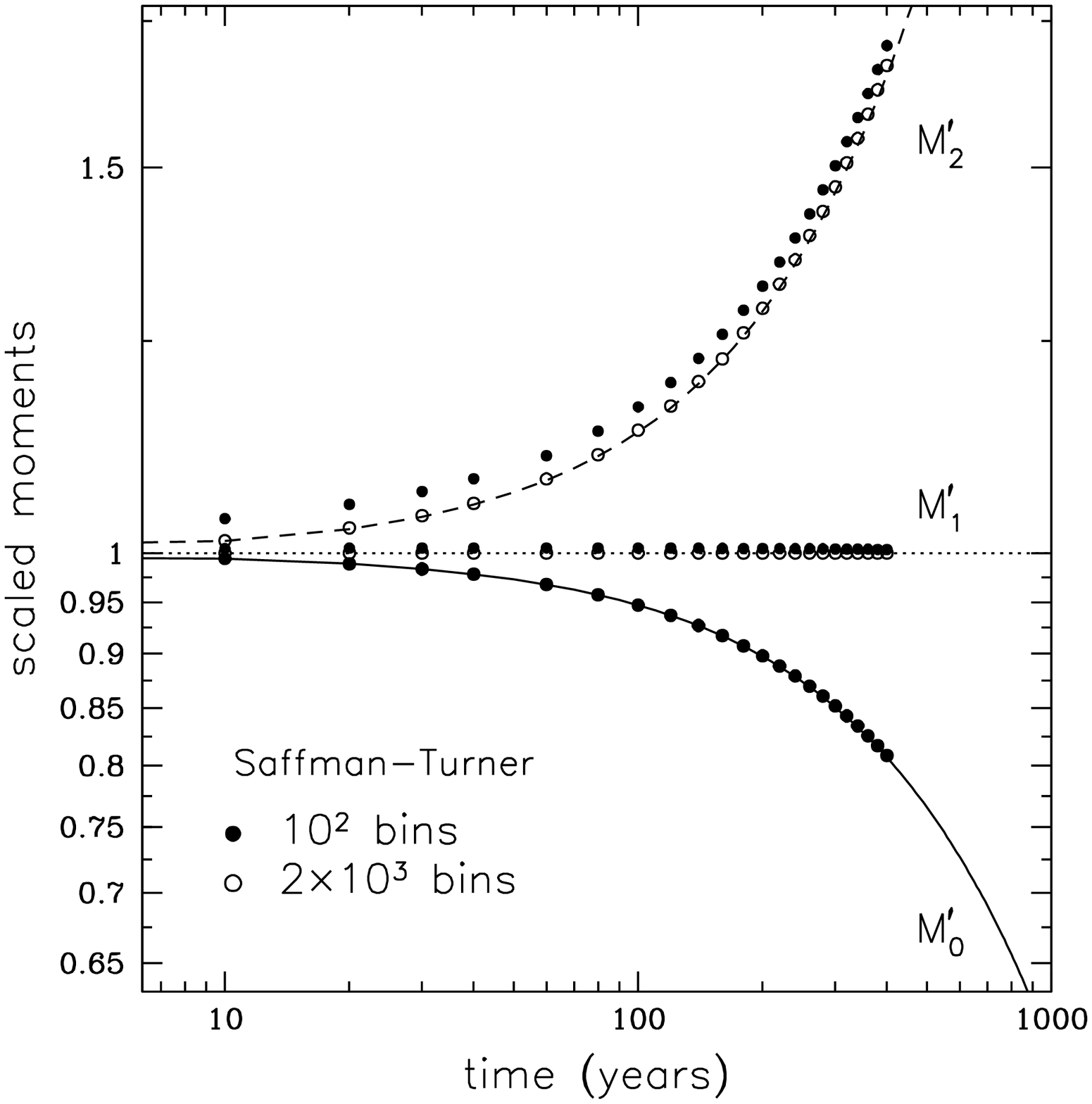}
\caption{Comparison between the scaled integer moments $M^\prime_k =
M_k/M_k(0)$ for $k = 0,1,2$ (curves) obtained
by the method of moments calculation with a 
simple Saffman-Turner turbulent coagulation kernel (Eq. 7) and a brute
force integration (symbols) of the coagulation equation (Eq. 1). The integer 
moments for the latter are calculated a priori using the distribution function 
$f(m,t)$. There is good agreement, especially for the higher resolution
case.}
\end{figure}

\clearpage

\begin{figure}
\plotone{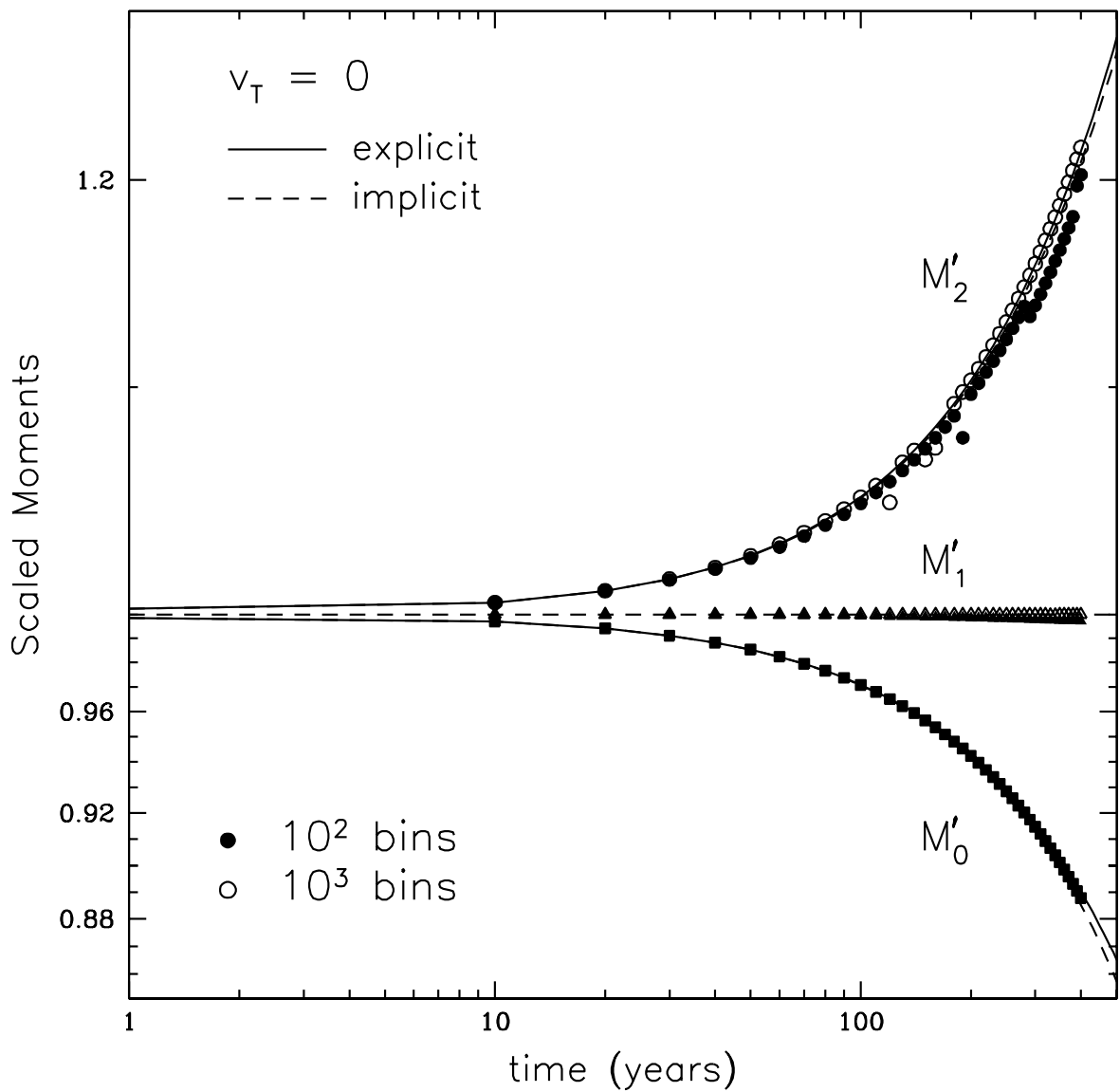}
\caption{Comparison of the scaled integer moments $k=0,1,2$ obtained from
the moments method (curves), and those obtained from the integration of the 
coagulation equation (Eq. 1, symbols) for the case of a realistic collision 
kernel with
$v_T = 0$. Both the explicit (solid curves), and implicit (short-dashed
curves) approaches match the coagulation calculation fairly well, especially
the higher resolution case.}
\end{figure}

\clearpage

\begin{figure}
\plotone{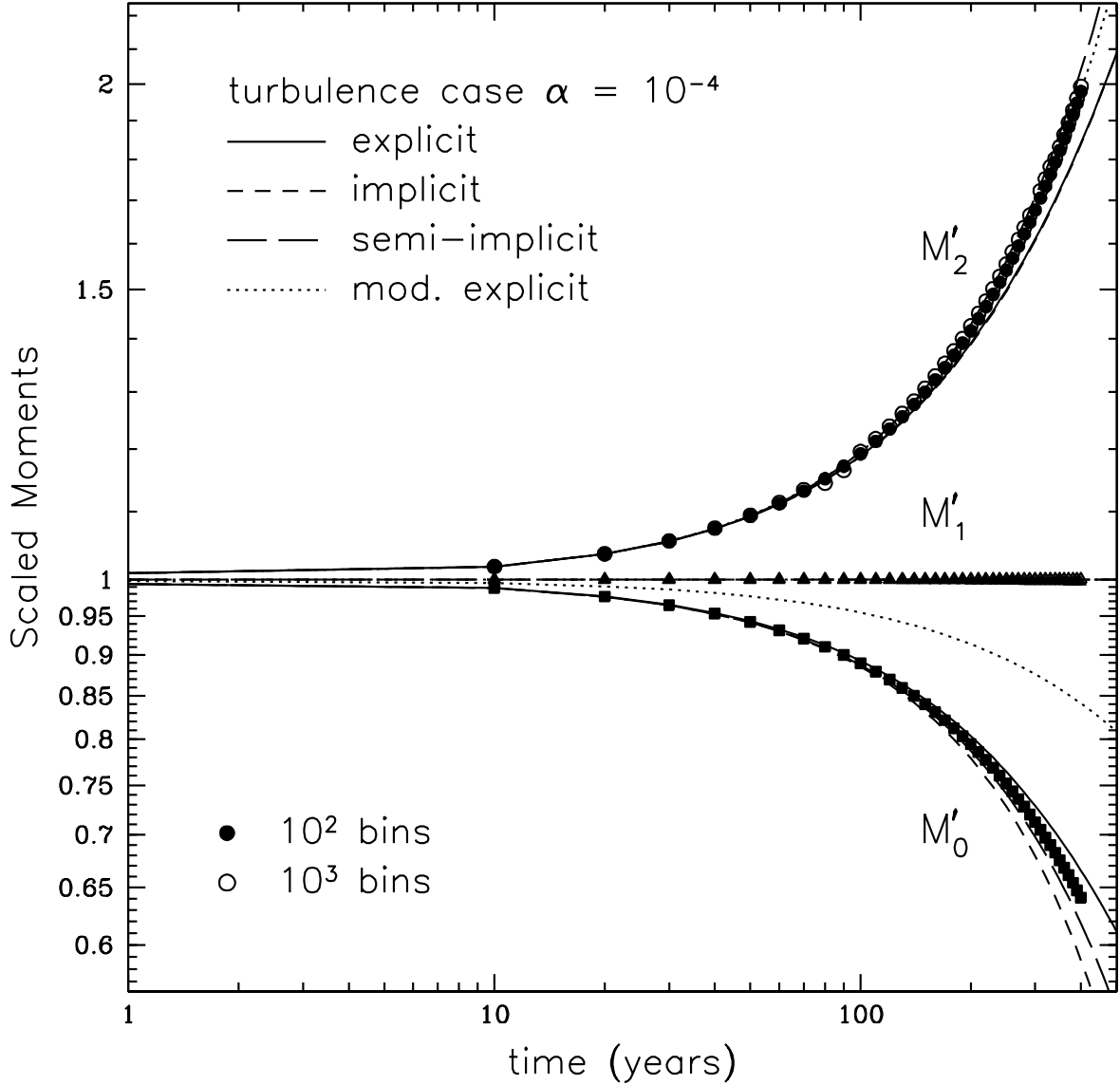}
\caption{Comparison of the scaled integer moments $M_k$ for $k=0,1,2$ obtained 
from the moments method, with results obtained from the brute force integration
of the coagulation
equation (Eq. 1) for the case of a realistic collision kernel with
$v_T \neq 0$ (and no systematic velocities). In this case the turbulence
parameter $\alpha = 10^{-4}$.  Both the explicit (solid curves) and
implicit (short-dashed curves) approaches have some difficulty matching 
the coagulation calculation for both $M_0$ and $M_2$ (note that the solid and 
short-dashed curves
for $M_2$ lie on top of each other). However, the semi-implicit approach
(long-dashed curves) provides a better fit. The modified explicit 
approach (dotted curves) provides a better fit for $M_2$ while giving a worse
fit for $M_0$ (see {\S} 3.2).}
\end{figure}

\clearpage

\begin{figure}
\plotone{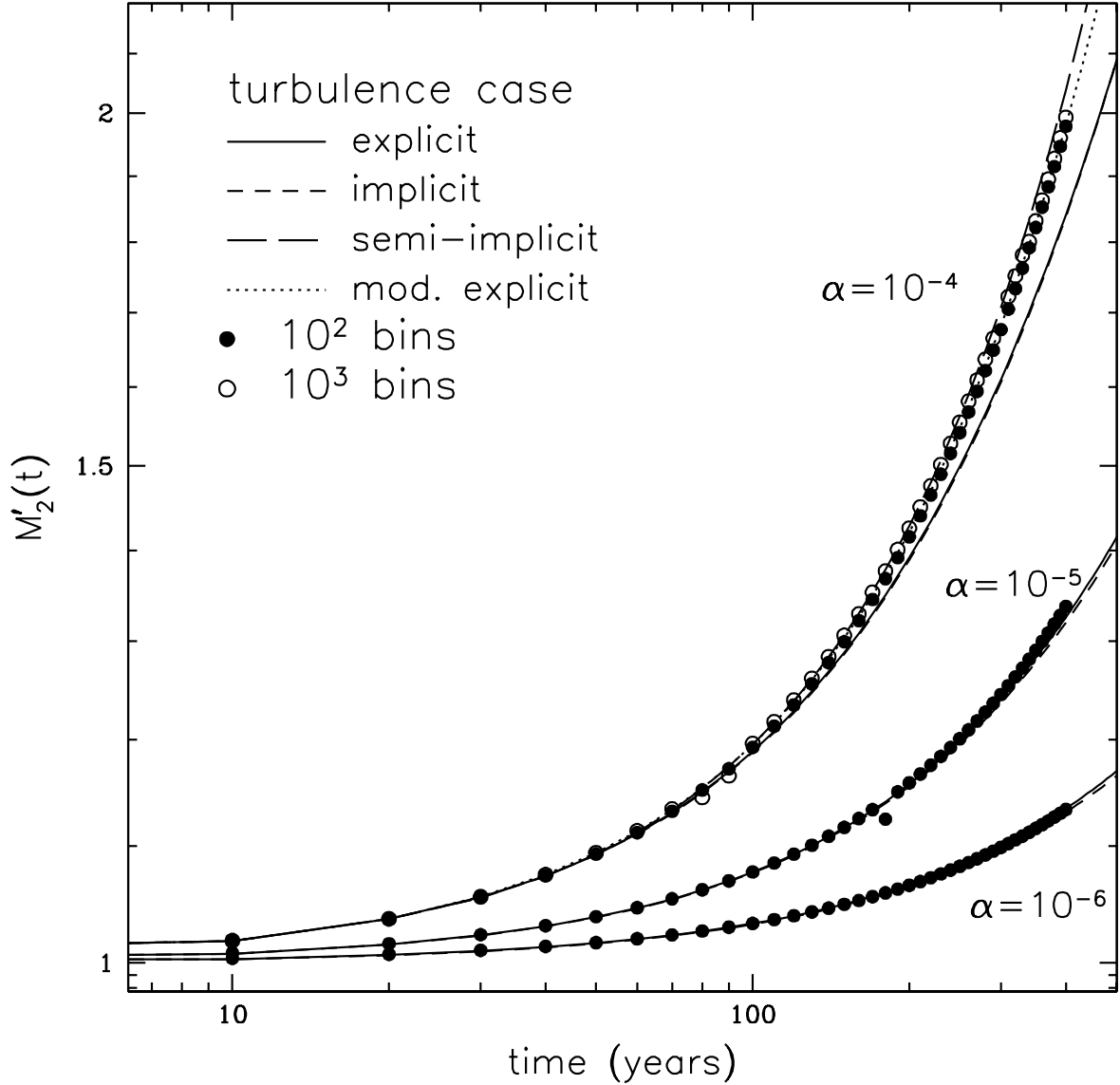}
\caption{Comparison of the second moment $M^\prime_2(t)$ obtained from the 
moments method with that obtained from integration of the coagulation 
equation (Eq. 1)
for the case of a realistic collision kernel with $v_T \neq 0$ (and no
systematic velocities) for three different values of the turbulence
parameter $\alpha$. Both the explicit (solid curves) and implicit
(short-dashed curves) approaches match the lower $\alpha$ values fairly well, 
but as
indicated in fig. 3, have difficulty matching the case of $\alpha = 10^{-4}$
(note that the solid and short-dashed curves lie on top of each other).
However, not surprisingly, the semi-implicit approach (long-dashed curves) 
and a modified explicit approach (for $M_2$ only, dotted curve) provides a
much better fit (see {\S} 3.2).}
\end{figure}

\clearpage

\begin{figure}
\plotone{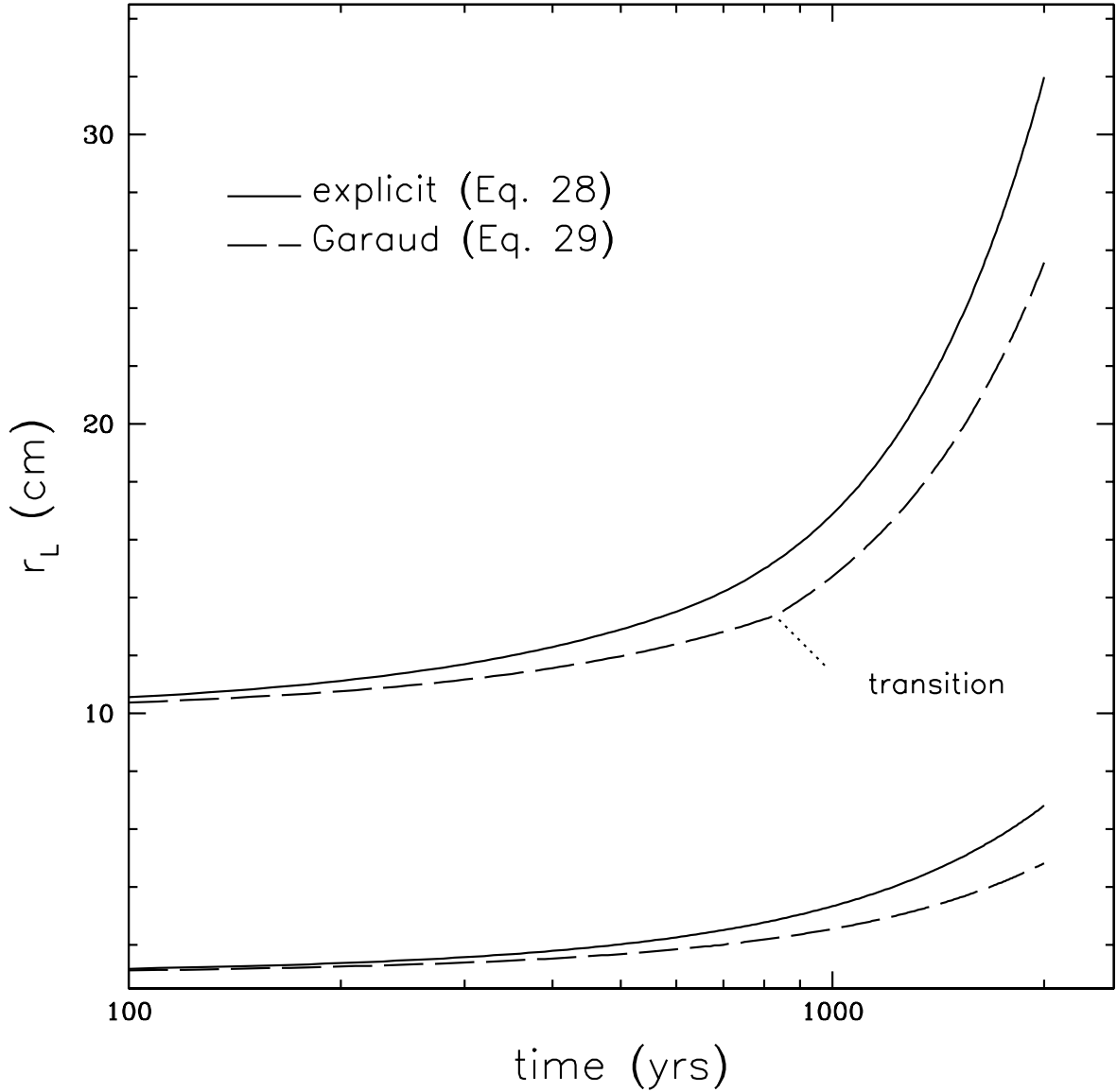}
\caption{Comparison of particle growth as a function of time using the
explicit approach (Eq. 28, solid curve), and the \citet{gar07} analytical
particle growth expression (long-dashed curve). The upper and
lower sets of curves
differ in the initial size of the largest particle. Both
sets of curves begin to diverge immediately. The somewhat subdued kink in the 
upper curves
at $r_L \sim 13.5$ cm is due to a shift from Epstein to Stokes flow.
However, overall, the agreement is not bad ($\sim 20-30$\% in $r_L$).}
\end{figure}

\clearpage

\begin{figure}
\plotone{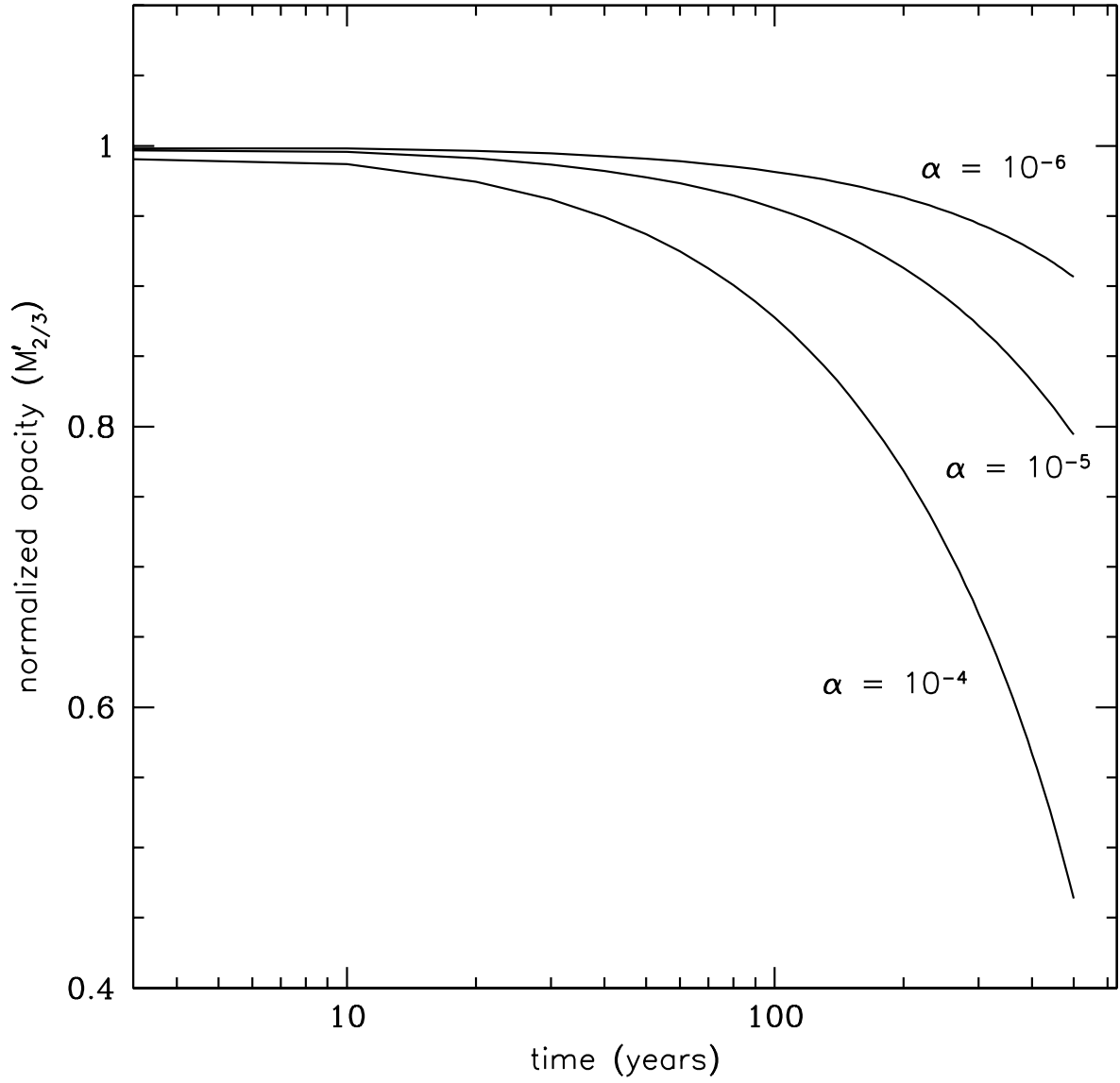}
\caption{Plot of the normalized (to initial value) wavelength-independent
opacities for different choices of the turbulence parameter $\alpha$. 
Opacities decrease sharply (in the absence of any mechanism to hinder particle
growth) for higher $\alpha$. Calculations were performed using the 
semi-implicit approach with both turbulent and systematic relative 
velocities included. }
\end{figure}


\begin{thebibliography}{}

\bibitem[Beckwith \& Sargent(1991)]{bec91} Beckwith, S. V. W., \& Sargent, 
A. I. 1991, \apj, 381, 250
\bibitem[Beckwith et al.(2000)]{bec00}  
Beckwith, S. V. W., Henning, T., \& Nakagawa, Y., In Protostars and Planets
IV, V. Mannings, A. P. Boss, and S. S. Russell, eds.,
(Tucson: Univ. of Arizona Press), 533
\bibitem[Brauer et al.(2008)]{bra08}
Brauer, F., Dullemond, C. P., \& Henning, Th., 2008, \aap, 480, 859
\bibitem[Bromley \& Kenyon(2006)]{bro06}
Bromley, B. C., \& Kenyon, S. J., 2006, \aj, 131, 2737
\bibitem[Cuzzi et al.(1993)]{cuz93}
Cuzzi, J. N., Dobrovolskis, A. R., \& Champney, J. M. 1993, \icarus, 106, 102
\bibitem[Cuzzi \& Hogan(2003)]{cuz03}
Cuzzi, J. N., \& Hogan, R. C. 2003, \icarus, 164, 127
\bibitem[Cuzzi \& Weidenschilling(2006)]{cuz06}
Cuzzi, J. N., \& Weidenschilling, S. J. 2006, In Metoerites of the Early
Solar System II, D. S. Loretta, and H. Y. McSween, Jr., eds., 
(Tucson: Univ of Arizona Press), 353
\bibitem[Dominik et al.(2007)]{dom07}
Dominik, C., Blum, J., Cuzzi, J. N., \& Wurm, G. 2007, In Protostars and 
Planets V, B. Reipurth, D. Jewitt, and K. Keil, eds., 
(Tucson: Univ. of Arizona Press), 783
\bibitem[Dullemond \& Dominik(2005)]{dul05}
Dullemond, C. P., \& Dominik, C. 2005, \aap, 434, 971
\bibitem[Durisen et al.(2007)]{dur07}
Durisen, R. H., Boss, A. P., Mayer, L., et al., 2007, 
In Protostars and Planets V, B. Reipurth, D. Jewitt, and K. Keil, eds, 
(Tucson: Univ of Arizona Press), 607
\bibitem[Garaud(2007)]{gar07}
Garaud, P, 2007, arXiv:0705.1563, Submitted to ApJ
\bibitem[Hubickyj et al.(2005)]{hub05}
Hubickyj, O., Bodenheimer, P., \& Lissauer, J. J., 2005, \icarus, 179, 415 
\bibitem[Johansen et al.(2007)]{joh07}
Johansen, A., Oishi, J. S., Low, M.-M. M., et al. 2007, \nat, 448, 1022
\bibitem[Kessler-Silacci et al.(2006)]{kes06}
Kessler-Silacci, J., Augereau, J.-C., Dullemond, C. P., et al. 2006, \apj,
639, 275
\bibitem[Kenyon \& Luu(1999)]{ken99}
Kenyon, S. J., \& Luu, J. X., 1999, \apj, 526, 465
\bibitem[Leinhardt \& Richardson(2005)]{lei05}
Leinhardt, Z. M., \& Richardson, D. C, 2005, \apj, 625, 427
\bibitem[Loginov(1979)]{log79}
Loginov, V. I. 1979, Dehydration and Desalinization of Oil: Khimiya, Moscow
\bibitem[Markiewicz et al.(1991)]{mar91}        
Markiewicz, W. J., Mizuno, H., \& V\"{o}lk, H. J. 1991, \aap, 242, 286
\bibitem[Marov \& Kolesnichenko(2001)]{mar01}
Marov, M. Ya., \& Kolesnichenko, A. V. 2001, Mechanics of Turbulence of
Multicomponent Gases: Kluwer Academic Publishers
\bibitem[Nakagawa et al.(1986)]{nak86}
Nakagawa, Y., Sekiya, M., \& Hayashi, C. 1986, \icarus, 67, 375
\bibitem[Natta et al.(2007)]{nat07}
Natta, A., Testi, L., Calvet, N., et al., 2007, In 
Protostars and Planets V , B. Reipurth, D. Jewitt, and K. Keil, eds, 
(Tucson: Univ of Arizona Press), 767
\bibitem[Ormel \& Cuzzi(2007)]{orm07}
Ormel, C. W., \& Cuzzi, J. N. 2007, \aap, 413, 466
\bibitem[Ormel et al.(2007)]{orm07b}
Ormel, C. W., Spaans, M., \& Tielens, A. G. G. M. 2007, \aap, 461, 215
\bibitem[Ormel et al.(2008)]{orm08}
Ormel, C. W., Cuzzi, J. N., \& Tielens, A. G. G. M. 2008, \aap, submitted
\bibitem[Pollack et al.(1996)]{pol96}
Pollack, J. B., Hubickyj, O., Bodenheimer, P., et al., 1996, \icarus, 124, 62
\bibitem[Press et al.(1992)]{pre92}
Press, W. H., Teukolsky, S. A., Vetterling, W. T., \& Flannery, B. P. 1992,
Numerical Recipes in Fortran: 2nd edition, Cambridge University Press.
\bibitem[Przygodda et al.(2003)]{pry03}
Przygodda, F., van Boekel, R., \`{A}brah\`{a}m, P., et al. 2003, \aap, 412,
L43
\bibitem[Saffman \& Turner(1956)]{saf56}
Saffman, P. G., \& Turner, J. S., 1956, J. Fluids Mech., 1, 16
\bibitem[Safronov(1969)]{saf69}
Safronov, V. S. 1969, Evolution of the Protoplanetary Cloud and the Formation
of the Earth and Planets: NASA TTF-677, 1972
\bibitem[Silk \& Takahashi(1979)]{sil79}
Silk, J., \& Takahashi, T., 1979, \apj, 229, 242
\bibitem[Smoluchowski(1916)]{smo16}
Smoluchowski, M., 1916, Phys. Z, 17, 557
\bibitem[Stepinski \& Valageas(1997)]{ste97}
Stepinski, T. F., \& Valageas, P., 1997, \aap, 319, 1007
\bibitem[Tanaka et al.(2005)]{tan05}
Tanaka, H., Himeno, Y., \& Ida, S., 2005, \apj, 625, 414
\bibitem[Trubnikov(1971)]{tru71}
Trubnikov, B. A. 1971, Soviet Phys. - Doklady, 16, 124
\bibitem[van Boekel et al.(2003)]{van03}
van Boekel, R., Waters, L. B. F. M., Dominik, C., et al. 2003, \aap, 400, L21
\bibitem[V\"{o}lk et al.(1980)]{vol80}
V\"{o}lk, H. J., Morfill, G. E., Roeser, S., \& Jones, F. C. 1980,
\aap, 85, 316
\bibitem[Weidenschilling(1977)]{wei77}
Weidenschilling, S. J., 1977, \mnras, 180, 57
\bibitem[Weidenschilling(1984)]{wei84}
Weidenschilling, S. J., 1984, \icarus, 60, 553
\bibitem[Weidenschilling(1997)]{wei97}
Weidenschilling, S. J., 1997, \icarus, 127, 290
\bibitem[Weidenschilling(2000)]{wei00}
Weidenschilling, S. J., 2000, \ssr, 92, 295
\bibitem[Weidenschilling(2002)]{wei02}
Weidenschilling, S. J., 2002, Met.\& Pla. Sci., 37, A148.
\bibitem[Weidenschilling(2004)]{wei04}
Weidenschilling, S. J., 2004, In Comets II, ed. M. C. Festou, H. U. Keller,
\& H. A. Weaver (Tucson: Univ. of Arizona Press), 97
\bibitem[Weidenschilling \& Cuzzi(1993)]{wei93}
Weidenschilling, S. J., \& Cuzzi, J. N. 1993, In Protostars and Planets
III, ed. E. H. Levy \& J. I. Lunine (Tucson: Univ. of Arizona Press), 1031
\end{thebibliography}
\end{document}